\documentclass[letterpaper,11pt]{article}
\usepackage{graphicx} 
\usepackage[paperwidth=8.5in, paperheight=11.80in,margin=1in]{geometry}
\usepackage{setspace}
\usepackage{tikz}
\usepackage[strict]{changepage}
\usepackage{amsthm}
\usepackage{float}
\usepackage{fancyhdr}
 \usepackage{setspace}
 \usepackage{amssymb}
 \usepackage{parskip}
 \usepackage{tabularx, booktabs}
 \usepackage{amsmath}
 \usepackage{graphicx}
\theoremstyle{definition}

\usepackage{authblk}
\usetikzlibrary{shapes, arrows.meta, positioning}
\usepackage{tcolorbox}
\usepackage{subcaption} 
\usepackage[utf8]{inputenc}
\usepackage{csquotes}

\usepackage[colorlinks=true, linkcolor=blue, citecolor=blue, urlcolor=blue]{hyperref}

\usepackage[
  backend=biber,
  bibstyle=apa,          
  citestyle=numeric-comp,
  sorting=none
]{biblatex}

\DeclareLanguageMapping{american}{american-apa}
\DeclareFieldFormat{labelnumberwidth}{\mkbibbrackets{#1}}
\defbibenvironment{bibliography}
  {\list
     {\printfield[labelnumberwidth]{labelnumber}}
     {\setlength{\labelwidth}{\labelnumberwidth}%
      \setlength{\leftmargin}{\labelwidth}%
      \setlength{\labelsep}{\biblabelsep}%
      \addtolength{\leftmargin}{\labelsep}%
      \setlength{\itemsep}{\bibitemsep}%
      \setlength{\parsep}{\bibparsep}}}
  {\endlist}
  {\item}

\begin{document}

\title{\textbf{Comparing Malaria Trends in the Comoros Islands: ARIMA Modeling and Retrospective Analysis}}
\author[1*]{Dennis Baidoo}
\author[1]{Emmanuel Kubuafor}
\author[1]{Okeke Onyedikachi Joshua}
\author[2]{Samuel Frimpong Osarfo}
\author[3]{Frank Amo Agyei-Owusu}
\author[4]{Justice Akuoko-Frimpong}
\author[5]{Nicholas Appiah}
\author[6]{Agnes Duah}
\author[1]{Robert Amevor}
\author[7]{Francis Effah Boateng}
\author[8]{Fuseini Aboagye}
\author[8]{Bright Benyi}

\affil[1]{Department of Mathematics and Statistics, University of New Mexico, NM, USA}
\affil[2]{Department of Statistics, Oklahoma State University, OK, USA}
\affil[3]{Department of Mathematical Sciences, University of Arkansas, AR, USA}
\affil[4]{Department of Biostatistics, University of Michigan, MI, USA}
\affil[5]{Department of Statistics and Mathematics, Texas Tech University, TX, USA}
\affil[6]{Department of Economics, Applied Statistics and International Business, New Mexico State University, NM, USA}
\affil[7]{School of Economics, University of Maine, Orono, ME, USA}
\affil[8]{Department of Statistics and Actuarial Science, University of Ghana, Accra}
\affil[ ]{*\texttt{Corresponding author: baidennwin@gmail.com}}
\date{}

\maketitle
\onehalfspacing
    
\section*{Abstract}

\textbf{Context}: Malaria remains a serious health challenge in the Comoros Islands, despite ongoing control efforts. Past studies have shown reductions in cases due to prevention and treatment measures, but little work has been done to forecast future malaria deaths and assess the long-term impact of these measures. \textbf{Methods}: Malaria mortality data from 1990 to 2019 were analyzed using an ARIMA(1,0,0) model. The model was validated through diagnostic tests, ensuring reliability for forecasting trends.
\textbf{Results}: The study confirmed significant reductions in malaria cases, such as in Grand Comoro, where cases fell from 235.36 to 5.47 per 1,000 people. The ARIMA model predicted that fatalities will remain low if current control measures, including bed nets, indoor spraying, and mass drug administration, are sustained. 
\textbf{Interpretation}: The findings highlight the success of these interventions in reducing malaria mortality. However, challenges like drug and insecticide resistance and financial limitations pose risks to further progress. Continued support and adaptation of strategies are essential to address these challenges and sustain low malaria mortality rates. \textbf{Conclusion}: The study demonstrates the effectiveness of malaria control efforts in the Comoros and underscores the importance of maintaining these measures to achieve malaria elimination and improve public health outcomes.

\textbf{Keywords}: Malaria fatalities, ARIMA modeling, Retrospective analysis, Time series analysis, Public health, Forecasting.

\newpage
\section{Introduction} 
Malaria continues to pose a significant public health issue in most sub-Saharan African countries, including Comoros, located in the Indian Ocean, with tremendously varying trajectories across the archipelago, despite the United Nations' Sustainable Development Goal 3 in ensuring healthy lives and promoting well-being for all walks of life \cite{worldhealthorganization_2020_front, unitednations_2024_goal,ouledi_1995_epidemiology}. However, current studies have shown a substantial reduction in malaria morbidity and mortality between 2010 and 2014, associated with an effective combination of preventive and control treatment measures \cite{saidabassekassim_2016_major}. Efforts to reduce malaria fatalities have included various interventions such as chloroquine (CQ), artemisinin-based combination therapies (ACT), insecticide-treated bed nets (ITNs), mass drug administration (MDA) as part of fast elimination of malaria by source eradication (FEMSE), and indoor residual spraying (IRS). These approaches have significantly contributed to lowering mortality rates associated with malaria \cite{saidabassekassim_2016_major,chakir_2017_control}. About 82\% of heads of households in the villages in Comoros knew that malaria is a transmissible disease, which made the MDA a success \cite{nadia_2023_knowledge}. Although significant progress has been made, challenges persist, including ACT-resistant malaria strains, mosquito pesticide resistance, and the need for continued financial support to achieve malaria elimination in the Comoros \cite{rebaudet_2010_genetic}. Recent studies on the \textit{Plasmodium falciparum} populations on Grande Comoros Island, one species of parasite that causes malaria in humans and is spread by the bites of female \textit{Anopheles mosquitoes},  have revealed a significant decrease in parasitic genetic diversity and various infections from 2006-2007 to 2013-2016, suggesting a tremendous decrease in malaria transmission intensity \cite{huang_2018_temporal,mcqueen_2010_comprehensive}.  As a result, it is crucial to consider the overall mortality cases from 1990 to 2019 to assess the long-term effectiveness of the interventions implemented in eradicating malaria in Comoros. Out of all the reviewed literature, there has not been a single paper forecasting the future burdens of malaria fatalities on the effectiveness of these treatment and preventive measures in this critical region. By analyzing the post-implementation of tools like IRS, ITNs, and the FEMSE strategy, this study will provide critical insights into the long-term impact of these interventions, aiding future efforts toward malaria elimination in the region. Building on previous research, this study evaluates the overall effectiveness of malaria control interventions in the Comoros by forecasting national-level malaria mortality trends, rather than examining trends for individual islands.

\section{Materials and Methods}
\subsection{Data Collection and Source}
The data used for the study are the global annual causes of death numbers for different countries [10]. Fatalities of malaria in Comoros were extracted from 1990 to 2019. The dataset was obtained from a reliable global source, ensuring accuracy and consistency in malaria fatality records for the Comoros Islands. The data analysis was conducted using RStudio version 2024.4.2.764 (Posit Software, PBC, Boston, MA). The Autoregressive Integrated Moving Average (ARIMA) time series model was employed to evaluate and predict trends in malaria fatalities across all islands. 

\subsection{Statistical Analysis and Model Selection}
A special case of the ARIMA model is the autoregressive with order p, AR(p), and it is formulated below as;

\begin{equation}
x_t = \phi_1 x_{t-1} + \phi_2 x_{t-2} + \cdots + \phi_p x_{t-p} + w_t,
\end{equation}

where $x_t$ is stationary (the current malaria fatalities at time $t$), 
$x_{t-i}$’s are the previous malaria fatalities for $i = 1, 2, \ldots, p$ lags, 
$w_t \sim wn(0,\sigma_w^2)$ is a white noise and $\phi_1, \phi_2, \ldots, \phi_p$ 
are constants (lags) with $\phi_p \ne 0$. If the mean, $\mu$, of $x_t$ is non-zero, 
we replace $x_t$ by $x_t - \mu$ in (1) and get

\begin{equation}
x_t - \mu = \phi_1 (x_{t-1} - \mu) + \phi_2 (x_{t-2} - \mu) + \cdots + \phi_p (x_{t-p} - \mu) + w_t
\end{equation}

\begin{equation}
x_t = \alpha + \phi_1 x_{t-1} + \phi_2 x_{t-2} + \cdots + \phi_p x_{t-p} + w_t
\end{equation}

where 
\[
\alpha = \mu (1 - \phi_1 - \cdots - \phi_p).
\]

The ARIMA(p,0,0) model is equivalent to an autoregressive with order p, AR(p) \cite{shumway_2025_time}. To assess model performance, we conducted several diagnostic checks. Residuals were tested for normality using the Shapiro-Wilk test, and heteroscedasticity was evaluated by plotting residuals against fitted values. The Ljung-Box test was used to check for autocorrelation, while the corrected Akaike Information Criterion (AICc) guided model selection. These diagnostics helped validate assumptions and ensure model robustness. ARIMA was chosen for its robustness in modeling and forecasting time series data with strong temporal dependencies. Since the Comoros malaria mortality data do not exhibit clear seasonal patterns, using a seasonality-based model is unnecessary. While machine learning models could capture complex nonlinear patterns, ARIMA provides a more interpretable and reliable approach for short-to medium-term forecasting. Its effectiveness in assessing long-term intervention impacts further supports its suitability for this study.

\section{Results of Analysis}

\subsection{Access to Malaria Prevention Measures and Malaria Case Distribution}
In 2019, access to malaria prevention measures, specifically LLINs, IRS, ACT, and FEMSE, varied across the studied African countries (Fig. 1, left). Botswana and Eswatini had the highest coverage, with a large proportion of their populations protected, whereas Eritrea, Comoros, Namibia, and South Africa had relatively lower coverage levels. The distribution of estimated malaria cases also varied significantly among these countries (Fig. 1, right). Eritrea and Comoros, with lower coverage rates, had a higher proportion of cases within their high-risk populations, while Botswana and South Africa showed lower malaria case shares, potentially reflecting the impact of higher prevention coverage. These findings underscore the relationship between malaria prevention access and case distribution across different at-risk groups.

\begin{figure}[htbp]
  \centering
  \begin{subfigure}{0.4\textwidth}
    \includegraphics[width=\linewidth]{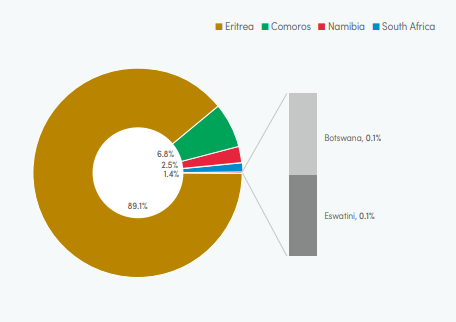}
  \end{subfigure} \hfill
  \begin{subfigure}{0.53\textwidth}
    \includegraphics[width=\linewidth]{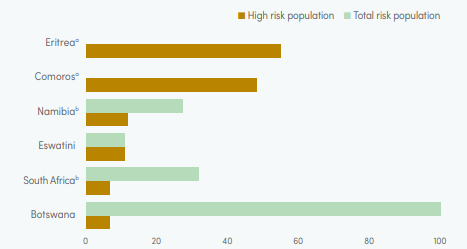}
  \end{subfigure}
  \caption{Percentage of population with access to prevention measures, 2019 (left) and share of estimated malaria cases, 2019 (right) (WHO, 2020)}
  \label{fig:AAA}
\end{figure}

Malaria cases by country in the WHO African Region for 2019 (Fig. 2, left), and reported indigenous malaria cases in countries with national elimination activities, comparing 2015 to 2019 (Fig. 2, right) (WHO, 2020). The bar chart on the left shows that malaria cases are highly concentrated in certain countries, with the highest burden observed in Nigeria, with Comoros among the lowest number of malaria cases recorded. The chart (Fig. 2, right) highlights changes in indigenous cases in countries pursuing elimination, showing reductions in South Africa, Botswana, and Eswatini, but increases in Namibia and Comoros over the period from 2015 to 2019.

\begin{figure}[htbp]
  \centering
  \begin{subfigure}{0.55\textwidth}
    \includegraphics[width=\linewidth]{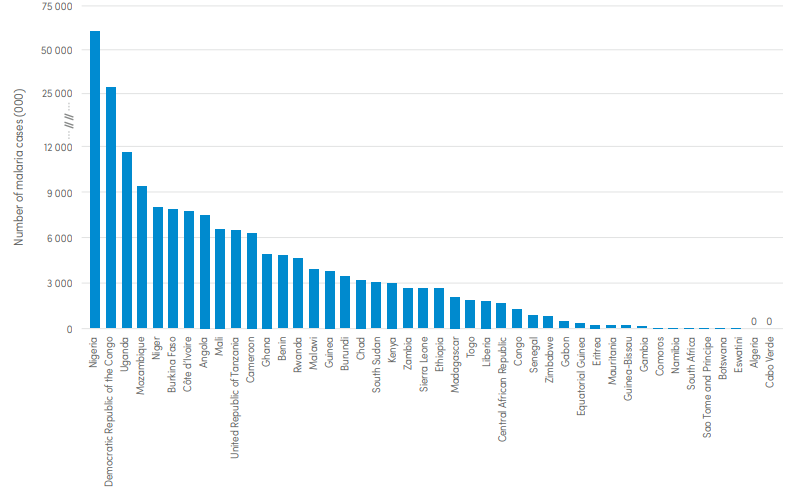}
  \end{subfigure}\hfill
  \begin{subfigure}{0.4\textwidth}
    \includegraphics[width=\linewidth]{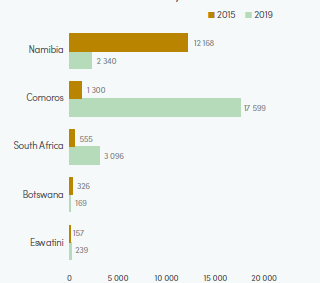}
  \end{subfigure}
  \caption{Malaria Cases by country in the WHO African Region, 2019 (left) and reported indigenous cases in countries with national elimination activities (right) (WHO, 2020)}
  \label{fig:BBB}
\end{figure}

\subsection{Model diagnostics and forecasting}
Both stepwise and search algorithms produced the same AICc (376.62). The best model generated by the ARIMA is ARIMA(1,0,0) with a mean term. The ARIMA(1,0,0) model’s residuals were evaluated for white noise, normality, and stationarity using the Box-Ljung test, Shapiro-Wilk test, and KPSS test, respectively. The ACF plot (Fig. 3, left) shows that autocorrelations fall within the confidence intervals, indicating no significant autocorrelation exists. Residual analysis (also in Fig. 3, left) suggests normality, with minimal deviations in mean and variance despite a slight dip in 2014. The tests confirm white noise (p = 0.834 $>$ 0.05), normality (p = 0.08205 $>$ 0.05), and stationarity (KPSS p = 0.1 $>$ 0.05). These diagnostics indicate that the ARIMA model is appropriate for forecasting the malaria burden in Comoros while accounting for evaluating the overall effectiveness of preventive intervention measures.

\begin{figure}[htbp]
  \centering
  \includegraphics[width=1\linewidth]{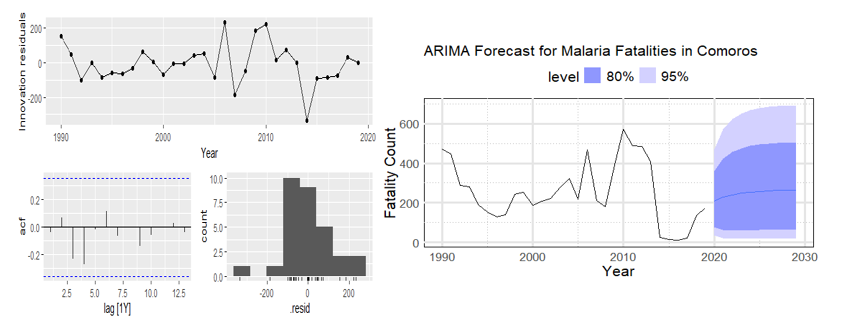}
  \caption{Residual analysis (left) and forecast of Malaria Fatalities in Comoros (right).}
  \label{fig:CCC}
\end{figure}

The Comoros Islands solely relied on chloroquine for malaria treatment and prevention after gaining independence. However, due to political instability, budget constraints, and chloroquine resistance, malaria cases continued to rise, as seen in the upward trend in the graph (Fig. 3, right) before 2004. The situation improved after 2003, when the government introduced artemether-lumefantrine and launched additional measures like indoor spraying, mosquito nets, and preventive therapy, which contributed to the decline in malaria fatalities shown in the graph (Fig. 3, right) after 2010, despite a slight increase in recent years \cite{silai_2007_surveillance,mukonoweshuroeliphasg_1990_the,a2016_african}. In the early 1990s, the exploration of how the genetic diversity and infection complexity of P. falciparum populations on Grande Comore Island shifted over time following the adoption of artemisinin-based combination therapy as the primary treatment for malaria. This had a significant impact on the rate of malaria fatalities across all islands and underserved regions \cite{huang_2018_temporal}.

The analysis emphasizes a clear link between malaria prevention coverage and case distribution: countries with higher preventive measure coverage tend to have lower malaria burden within high-risk populations, suggesting that expanding access to these measures may reduce cases. Additionally, the diagnostic validation of the ARIMA model reinforces its potential as a reliable tool for forecasting malaria trends in regions with limited data, such as Comoros, enabling data-driven planning for malaria interventions.

\section{Discussion}
The primary aim of this article is to examine trends in malaria fatalities across all Comoros islands from 1990 to 2019 and help evaluate the treatment and prevention measures implemented to curb the cases by forecasting future malaria burdens. The findings from the recent pieces of literature provide valuable insights into the trends of malaria morbidity and mortality in the Comoros, highlighting the effectiveness of the implemented control measures. The studies' retrospective descriptive analysis and trend assessment offer a comprehensive understanding of the malaria situation in the Comoros during the specified period. In contrast, the ARIMA modeling focuses on forecasting the future burden of malaria in Comoros based on historical data from 1990 to 2019. The ARIMA approach takes into account the temporal dependencies and patterns in the malaria fatalities data, allowing for more accurate predictions of future malaria fatalities. The model’s satisfactory performance, as indicated by the residual diagnostics, strengthens the reliability of the forecasts (Fig. 3). The combination of the retrospective analysis and the ARIMA-based forecasting approach provides a more comprehensive understanding of the malaria situation in Comoros.

 Malaria is responsible for 15-30\% of hospital admissions and 15-20\% of child deaths in the region. Transmission rates vary across the islands, with Anopheles gambiae and Anopheles funestus being the main mosquito vectors, and P. falciparum causing the majority of cases. The islands experience high endemicity with stable transmission, although conditions differ both between and within the islands. Control measures have included national policies, vector control strategies, and treatment guidelines. Recently, mass treatment with artemisinin-based combination therapy has shown promising results on two of the islands \cite{ouledi_1995_epidemiology}. 
 
Malaria fatalities in Comoros showed significant fluctuations from 1990 to 2010, with a major peak around 2010 reaching approximately 600 deaths per year, followed by a sharp decline after 2015. This drop coincided with the introduction of artemisinin-based combination therapies, indoor residual spraying, and insecticide-treated bed nets. Fatalities stabilized around 100–200 deaths per year by 2020, reflecting the sustained impact of these interventions. The ARIMA forecast (2020–2030) suggests that if control measures continue, fatalities may remain low, but without sustained efforts, cases could rise again, as indicated by the widening confidence intervals. This highlights the importance of ongoing prevention strategies to prevent a resurgence in malaria deaths.

While the World Health Organization (WHO) demonstrates the effectiveness of control measures in reducing the malaria burden, the ARIMA model offers insights into the potential future trajectory of malaria fatalities, aiding in resource allocation and interventional planning. From 2010 to 2014, the Comoros Islands experienced a substantial decline in malaria morbidity and mortality, attributable to the implementation of comprehensive prevention and control measures \cite{saidabassekassim_2016_major}. In addition, the forecast (Fig. 3, right) suggests that with continuous implementation of these treatment and prevention measures, there is a high chance of total elimination of malaria morbidity and mortality from these regions. While the ARIMA model provides reliable forecasts based on historical trends, it does not explicitly account for emerging challenges such as ACT-resistant malaria strains and mosquito pesticide resistance. These factors could alter transmission dynamics and intervention effectiveness, potentially leading to higher-than-predicted malaria fatalities if resistance spreads and reduces the efficacy of current control measures. However, it is important to note that the ARIMA model’s forecasts are based on prior patterns and do not explicitly account for the impact of specific control measures or changes in health policies. Therefore, the malaria forecast is the overall evaluation of the effectiveness of all implemented preventive and treatment measures to curb the associated fatalities in conjunction with the findings of other relevant studies to obtain a holistic view of the malaria situation in Comoros.

\section{Conclusion}
Recent studies and analyses using ARIMA modeling provide complementary perspectives on the malaria situation in Comoros. While these studies highlight the significant decline in malaria morbidity and mortality due to effective control measures, the ARIMA model offers insights into the future burden of malaria based on historical patterns. The combination of these analyses emphasizes the importance of sustained efforts in malaria prevention, control, and surveillance to maintain the progress achieved and further reduce the burden of malaria in Comoros. Continued research investment, new tools and strategies development, and healthcare system strengthening will be crucial in the ongoing fight against malaria in Comoros. Regular monitoring, evaluation, and adaptation of control measures based on the changing epidemiology and forecast trends will be essential to optimize the allocation of resources and achieve the ultimate goal of malaria elimination in the Comoros Islands \cite{worldhealthorganization_2020_front,unitednations_2024_goal,saidabassekassim_2016_major}.

This study contributes to Sustainable Development Goal (SDG) 3: "Good Health and Well-being" by highlighting the importance of equitable access to malaria prevention measures, such as long-lasting insecticidal nets, indoor residual spraying, and antimalarial treatments like ACT. By demonstrating that countries with higher prevention coverage experience a lower malaria burden, the study underscores the need for targeted interventions in regions with limited access to preventive resources. Despite progress, challenges remain. Ongoing research and customized strategies are essential for successfully controlling malaria in the Comoros \cite{chakir_2017_control}. Expanding these measures can reduce malaria cases and fatalities, thereby improving health outcomes in vulnerable communities and advancing progress toward SDG 3.

To improve malaria prevention and reduce disparities, efforts should focus on making insecticide-treated bed nets and indoor spraying more accessible, especially in underserved areas. Better tracking of malaria cases can help target high-risk regions and use resources wisely. Rotating insecticides and monitoring drug effectiveness can help prevent resistance. More funding from governments and global organizations is needed to keep these efforts going. Simple health campaigns can teach people about using bed nets and seeking treatment early. Expanding healthcare services, like mobile clinics and better drug supply, will ensure more people get diagnosed and treated. Combining malaria control with other health programs can also make interventions more effective. These steps can help reduce malaria cases and move closer to elimination.

\section*{Acknowledgment}
\vspace{-0.5cm}
The corresponding author extends sincere thanks to all co-authors for their partnership, insightful input, and steadfast support that greatly enriched this research.
\vspace{-0.5cm}
\section*{Conflict of Interest}
\vspace{-0.5cm}
The authors declare no conflict of interest.
\vspace{-0.5cm}

\section*{Funding}
\vspace{-0.5cm}
No specific funding was received from public, commercial, or not-for-profit organizations for this research.

\newpage

\printbibliography

\end{document}